\documentclass[12pt,preprint]{aastex}

\def\sun{\odot}
\def\earth{\oplus}
\begin{document}
\title{The effect of Semi-Collisional Accretion on Planetary Spins}
\shortauthors{Schlichting & Sari}
\shorttitle{The spin of terrestrial planets}
\author{Hilke E. Schlichting and Re'em Sari}
\affil{California Institute of Technology, MC 130-33, Pasadena, CA 91125} 
\email{hes@astro.caltech.edu, sari@tapir.caltech.edu}
\slugcomment{Submitted to ApJ}

\begin{abstract}
Planetesimal accretion during planet formation is usually treated as
collisionless. Such accretion from a uniform and dynamically cold disk
predicts protoplanets with slow retrograde rotation. However, if the building
blocks of protoplanets, planetesimals, are small, of order of a meter in size,
then they are likely to collide within the protoplanet's sphere of
gravitational influence, creating a prograde accretion disk around the
protoplanet. The accretion of such a disk results in the formation of
protoplanets spinning in the prograde sense with the maximal spin rate allowed
before centrifugal forces break them apart. As a result of semi-collisional
accretion, the final spin of a planet after giant impacts is not completely
random but is biased toward prograde rotation. The eventual accretion of the
remaining planetesimals in the post giant-impact phase might again be in the
semi-collisional regime and delivers a significant amount of additional
prograde angular momentum to the terrestrial planets. We
suggest that in our Solar System, semi-collisional accretion gave rise to the
preference for prograde rotation observed in the terrestrial planets and
perhaps the largest asteroids.
\end{abstract}

\keywords{planets and satellites: formation --- solar system: formation}

\section{INTRODUCTION}
Protoplanets form by the accretion of planetesimals. When planetesimals are
accreted, they deliver rotational angular momentum due to their relative
motion with respect to the protoplanet. This accretion is usually treated as
collisionless, assuming that collisions among planetesimals can be neglected
while they are within the Hill sphere of the protoplanet. In collisionless
accretion, the angular momentum accreted from a uniform and dynamically cold
disk of planetesimals results in slow retrograde rotation
\citep{LK91,DT93}. \cite{LBGK97} have shown that rapid prograde rotation can
only be achieved if disk density profiles are imposed such that the surface
mass density near the outer edges of a protoplanet's feeding zone is
significantly greater than that in the rest of the accretion zone. This
suggests that protoplanets do not possess any significant spin due to
collisionless planetesimal accretion. The final stage of terrestrial planet
formation consists of collision and accretion events of a few dozen
protoplanets of about $0.05~M_{\earth}$ \citep{ACL99,C01,GLS04}. These giant
impacts deliver spin angular momentum to the final planet. If giant impacts
are solely responsible for the final spin properties of terrestrial planets
then terrestial planets should display random obliquities (the angle between
the orbital and rotational angular momentum) and exercise prograde and
retrograde rotation with equal likelihood.

In this paper, we investigate the possibility of semi-collisional or
collisional planetesimal accretion and the effect it would have on planetary
spins. In \S 2 we first determine the range of planetesimal sizes for which
semi-collisional or collisional accretion applies and derive the consequences
of semi-collisional and collisional accretion for the spin of
protoplanets. The spin of terrestial planets due to giant impacts of
protoplanets is calculates in \S 3 and compared with the semi-collisional
contribution. Post giant-impact accretion is discussed in \S 4.  Comparison
with the Solar system and conclusions follow in \S 5.

\section{SEMI-COLLISIONAL AND COLLISIONAL ACCRETION} 
\begin{figure}
\plotone{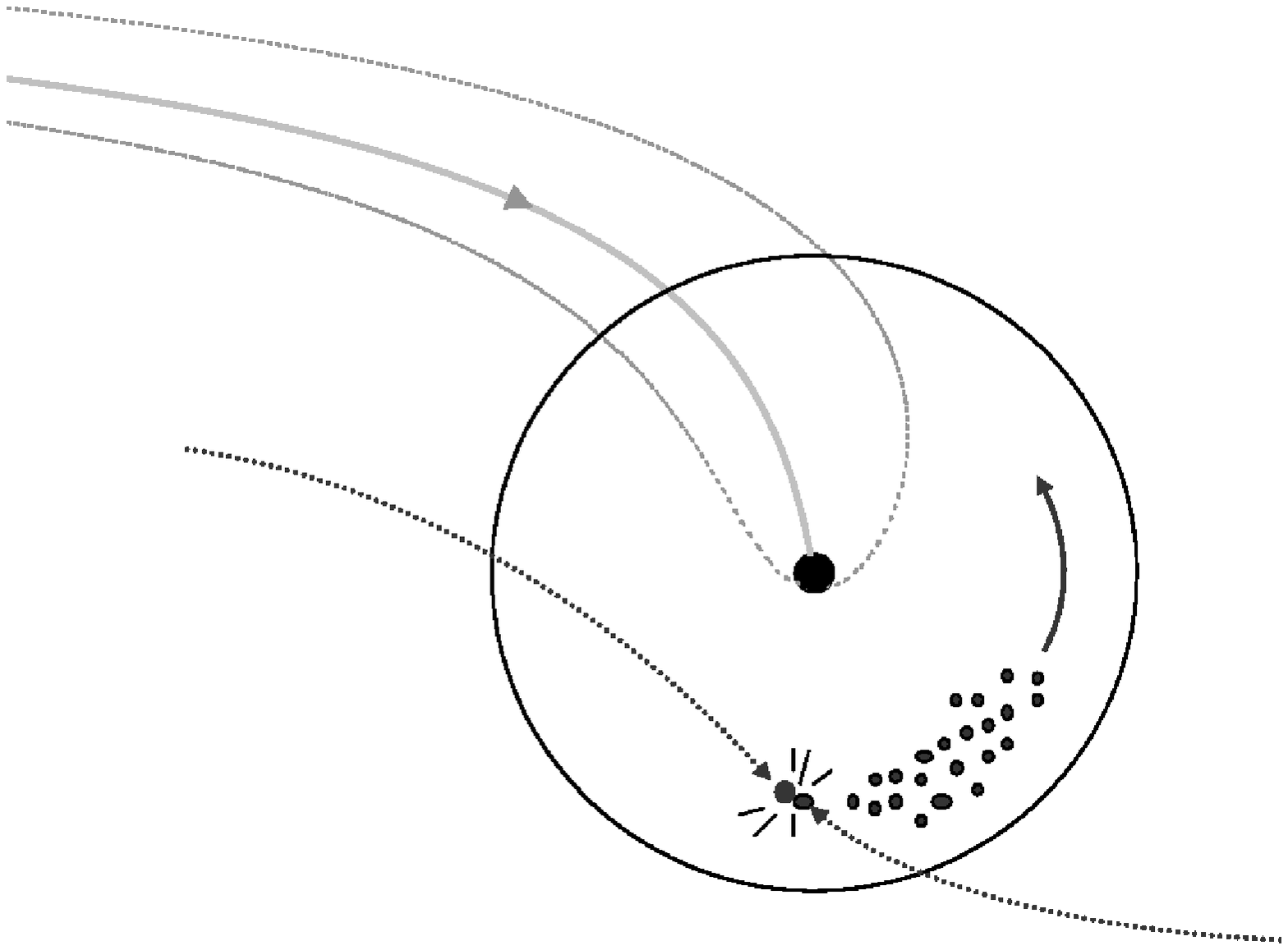}
\caption{ Collisionless and semi-collisional accretion. The protoplanet is
represented by the filled black circle and its Hill radius is given by the
solid black line. In the case of collisionless accretion (light grey) only
planetesimals with impact parameters that allow direct collision with the
protoplanet are accreted. In the semi-collisional and collisional regimes
(dark grey) unbound planetesimals collide inside the Hill sphere of the
protoplanet producing bound planetesimals which form a prograde accretion disk
around the protoplanet. This enables the protoplanet to effectively accrete at
its Hill radius. 
 }
\label{fig3}
\end{figure}

The Hill radius
denotes the distance from the protoplanet at which the tidal force due to the
Sun and the gravitational force due to the protoplanet both acting
on a planetesimal are in equilibrium. It is given by
\begin{equation}
R_{H} \equiv a \left( \frac{m}{3 M_{\sun}}\right) ^{1/3} 
\end{equation}
where $a$ is the semi-major axis of the protoplanet and $m$ its mass. When two
planetesimals collide with each other while passing through the Hill sphere of
the protoplanet, one or both of them become bound to the protoplanet. Further
collisions among the bound particles damp their random motions, leading to the
formation of an accretion disk around the protoplanet \citep{SG06} (see figure
\ref{fig3}). Inelastic planetesimal collisions and subsequent capture by the
planet's gravitational field has been proposed in order to form
circumplanetary disks from which regular satellites could form
\citep[e.g.][]{SPRV89,EM06}. Here we are exploring the possibility that the
growth of protoplanets is dominated by the accretion of such a planetesimal
disk. The details of this accretion process, such as what fraction of bound
particles will be accreted by the growing protoplanet, are
uncertain. Perturbations from nearby protoplanets and moons or gas, if still
present at the time protoplanets form, may facilitate the dissipation of the
planetesimals' angular momentum allowing efficient accretion onto the
protoplanet.

\subsection{Planetesimal Sizes}
For dynamically cold planetesimal disks, the ratio between the rate of
planetesimal collisions within the Hill sphere to the rate of direct
collisions onto the protoplanet is $\tau_g \alpha^{-1/2}$, where $\tau_g$ is
the optical depth within the disk plane over a distance of $R_H$ and $\alpha
\equiv r/R_H$ where $r$ is the protoplanet's radius. If $\tau_g \alpha^{-1/2}
> 1$, the accretion may be dominated by binding planetesimals into an
accretion disk rather than direct impacts onto the protoplanet; we call this
semi-collisional accretion. Collisional accretion takes over for
$\tau_g>1$. For inelastic planetesimals with velocity $u < v_{H} \equiv \Omega
R_{H}$, the optical depth in the disk is given by $\tau_g \sim 3 \sigma v_H /
s\rho_s u$ where $\Omega$, $s$, $\rho_s$ and $\sigma$ are the protoplanet's
Keplerian angular velocity around the Sun, the typical planetesimal radius,
material density and overall mass surface density respectively. The random
velocities of the planetesimals are damped by mutual collisions and stirred by
gravitational interactions with the protoplanets. When these two processes are
in equilibrium, we have
\begin{equation}\label{e1}
\frac{u}{v_{H}} \sim \alpha^{-2} \frac{\Sigma}{\sigma}
\frac{s}{r},~~~~~~~~~~~~~~{\rm for}~ u < v_{H} 
\end{equation}
where $\Sigma$ is the mass surface density of the protoplanets
\citep{GLS04}. Most of the planetesimal accretion occurs when $\Sigma / \sigma
\sim 1$. The condition for semi-collisional accretion ($\tau_g \alpha^{-1/2}>
1$) together with equation \ref{e1}, defines an upper limit to the
planetesimal size for which semi-collisional accretion holds. Using the
minimum mass solar nebula \citep{H81} surface density of $ \sim 8~{\rm
g/cm^2}$ at $1~{\rm AU}$, $ \rho_s \sim 3~{\rm g/cm^3}$ and an isolation mass
$ \sim 0.05~M_{\earth}$ \citep{WSDMO97}, we find \footnote{All estimates above
assumed $u<\alpha^{1/2} v_{H}$. However, for large enough planetesimals we
have $\alpha^{1/2} v_{H} < u < v_{H}$. Taking this into account results in a
slightly higher upper limit of $17~{\rm m}$ for $s$. For simplicity and given
the order of magnitude nature of this calculation, we ignore this
complication.} that $s\lesssim9~{\rm m}$. A lower limit to the planetesimal
size is given by the velocity dispersion for which the disk becomes locally
unstable to gravitational collapse. This velocity is $ \sim 10~{\rm cm/s}$ at
$1~{\rm AU}$, corresponding to a minimum size for planetesimals of $ \sim
6~{\rm cm}$. Therefore semi-collisional or collisional accretion applies as
long as $6~{\rm cm} \lesssim s \lesssim 9~{\rm m}$.  A fragmentation cascade
produced by destructive planetesimal collisions leads to the formation of ever
smaller planetesimals \citep{GLS04}. In fact, gravitational instabilities in
the disk may be responsible for the lower limit on the planetesimal size, in
which case $s\sim 6~$cm. Possible gaseous remnants of the solar nebula may
lower the velocity dispersion preventing fragmentation down to the stability
limit. Though this is an uncertainty during protoplanet formation it is
unlikely that significant amounts of gas prevailed after giant impacts.
Further, the low bulk density ($\sim 0.6~{\rm g/cm^3}$) of comets
\citep{A05,DG06} seems to suggest gentle accretion of small bodies and
therefore supports the idea of semi-collisional or collisional accretion.

\subsection{Spin of Protoplanets due to Planetesimal Accretion}
\begin{figure}
\plotone{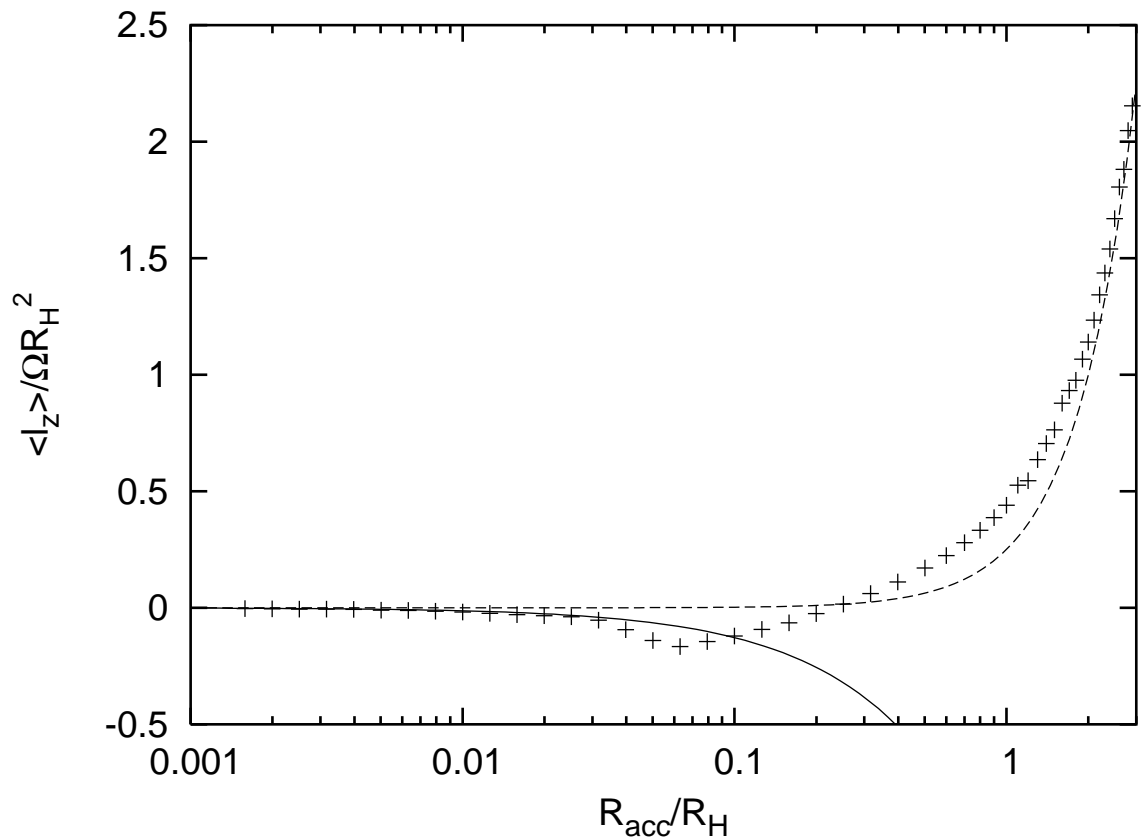}
\caption{Mean specific angular momentum $<l_z>$ in units of $\Omega R_{H}^2$
  accreted from a cold disk of planetesimals vs accretion radius $R_{acc}$ in
  Hill radii. The crosses indicate the results from our numerical integration
  and the dashed line corresponds to the limit in which the gravity of the
  protoplanet can be neglected, i.e. $R_{acc} >> R_{H}$. The solid line shows
  the analytic solution valid for $R_{acc} << R_H$ \citep{DT93}.
}
\label{fig1}
\end{figure}

We assume that the orbits of the planetesimals and the protoplanets are
circular and co-planar. The interaction between the planetesimals and the
protoplanet can be described by Hill's equations \citep{H78,GT80,PH86}. In our
coordinates the position of the planetesimal is given with respect to the
protoplanet. The $x$-axis points radially outwards and the $y$-axis in the
prograde direction. The equations of motion are given by
\begin{equation}\label{e2}
\ddot{x}-2\Omega \dot{y} -3 \Omega^2 x=-\frac{G m}{(x^2+y^2)^{3/2}} x 
\end{equation}
\begin{equation}\label{e3}
\ddot{y}+2\Omega \dot{x}=-\frac{G m}{(x^2+y^2)^{3/2}} y.
\end{equation} 
We solve these equations numerically and sum the specific angular momentum of
all planetesimals that pass within some effective accretion radius
$R_{acc}$. In collisionless accretion the protoplanet accretes at its actual
radius so that $R_{acc} = r$; in semi-collisional or collisional accretion an
accretion disk forms and the protoplanet effectively accretes at its
gravitational radius such that $R_{acc} \sim R_H$.  Figure \ref{fig1} shows
that protoplanets acquire a retrograde spin for $R_{acc} < 0.2~R_{H}$ and a
prograde rotation for $R_{acc} \sim R_H$. The prograde rotation for $R_{acc}
>> R_H$ can be understood by considering the angular momentum supplied by
planetesimals solely due to the Keplerian shear of the disk. In this case the
specific angular momentum acquired by the protoplanet is $ R_{acc}^2 \Omega/4
$ in the prograde sense \citep{LK91,DT93}. The actual angular momentum
delivered to the planet is given by figure \ref{fig1} for collisionless
accretion only. In the semi-collisional and collisional cases, the disk must
lose angular momentum before it can be accreted by the protoplanet. The
accretion of such a disk results in the formation of protoplanets spinning in
the prograde sense with the maximal spin rate allowed before centrifugal
forces break them apart.

\section{GIANT IMPACTS}
The final stage of terrestrial planet formation consists of collision and
accretion events among the protoplanets. These giant impacts deliver spin
angular momentum to the final planet. Provided that the random velocities of
the protoplanets are sufficiently large, one can neglect the shear imposed by
the differential rotation of the disk, so there is no preferred direction for
giant impacts to occur. Giant impacts therefore deliver angular momentum in a
random walk like fashion. \citet{LS91} and \citet{DT932} calculated the
magnitude of the random component of the spin angular momentum due to a single
giant impact and compared it with the observations. Here we determine the random
and systematic spin angular momentum delivered to the final planet by $N$
giant impacts using the following toy model. We start with $N+1$ identical
protoplanets all of mass $m$ and radius $r$ which are sequentially accreted
one by one. After $N$ such accretion events, we are left with a final planet
of mass $M=(N+1)m$ and radius $R=(N+1)^{1/3}r$. We assume throughout that
protoplanets are spherical with constant density $\rho$.

\subsection{Random Component of the Angular Momentum}
In the limit that the protoplanets' velocity dispersion is small compared to
their impact velocity and assuming that protoplanets have no spin, the maximum
angular momentum delivered by one impact is
\begin{equation}
l_{max} = \frac{M_T m}{M_T+m} \sqrt{ 2G \left( M_T+m \right) \left( R_T+r \right) }
\end{equation}
where $M_T$ is the mass and $R_T$ the radius of the target. The root mean
square (rms) angular momentum in the direction perpendicular to the plane of
the Solar System ($z$-direction) contributed by a single impact can be
obtained by averaging over all possible impact parameters and is given by
$l^z_{rms}=\sqrt{1/6} l_{max}$. Adding the contributions of each impact in
quadrature, with $M_T=nm$ and $R_T=n^{1/3}r$ for $n=1,2,...,N$, the final rms
angular momentum in the $z$-direction after $N \gg 1$ impacts is
\begin{equation}\label{e22}
L^z_{rms} =  
\sqrt{  {1 \over 7} } N^{-1/2} \omega_{crit} M R^2
\end{equation}
where 
\begin{equation}
\omega_{crit} = \sqrt{ \frac{4 \pi G \rho}{3} }.
\end{equation}
The precise number of giant impacts during the late stage of planet formation
is uncertain. However, the final ``isolation'' mass for the minimum mass solar
nebula at $1~{\rm AU}$ is about $0.05~M_\earth$ \citep{WSDMO97,GLS042}. This
suggests that about $20$ giant impacts have occurred in order to form an earth
at $1~{\rm AU}$. For $N \sim 20$, equation \ref{e22} predicts a spin period of
$\sim 4~{\rm hours}$ for the Earth. N-body simulations find a somewhat shorter
spin period of $\sim 1.8~{\rm hours}$ for bodies more massive than
$0.5~M_{\earth}$ \citep{ACL99}. This rapid rotation originates from unphysical
mergers between protoplanets encountering each other at more than the escape
velocity. As expected, N-body simulations also show that final obliquities due
to giant impacts with no initial spin are randomly distributed
\citep{ACL99,C01}.

\subsection{Systematic Component of the Angular Momentum}
The final spin of a terrestrial planet after giant impacts is no longer random
but contains a systematic component if each protoplanet possess a systematic
spin due to semi-collisional planetesimal accretion. The systematic component
of the angular momentum delivered by $N$ impacts of maximally spinning
protoplanets with prograde rotation is
\begin{equation}\label{e8}
L_{Spin} = L^z_{Spin} = {2 \over 5} N^{-2/3} M R^{2} \omega_{crit}.
\end{equation}

\subsection{Comparison}
Comparing the random  $z$-component of the angular momentum (equation
\ref{e22}) to the systematic one (equation \ref{e8}), we find that they are
similar in magnitude with the random component up to twice the systematic one
for $1 \lesssim N \lesssim 60$. The final distribution for the $z$-component
of the angular momentum is obtained by combining the random and the ordered
contributions. It is normally distributed with its mean given by equation
\ref{e8} and its standard deviation given by equation \ref{e22}. Since the
mean is positive, corresponding to prograde rotation, we expect more prograde
than retrograde spins in a given planetary system. We find about 70\% of all
planets to be rotating in the prograde sense and only 30\% in a retrograde
manner for $10 \lesssim N \lesssim 60$ giant impacts.

\subsection{Uncertainties}
The following uncertainties could affect our estimates for prograde and
retrograde rotation. We have assumed that the velocity dispersion of the
protoplanets is small compared to the impact velocity. However, the velocity
dispersion might be as large as the escape velocity from the protoplanet, in
which case the random component of the angular momentum could increase up to
$\sim \sqrt{2}$. A higher fraction of planets with retrograde rotation would
be produced if the mutual accretion of protoplanets were pairwise, such that
all giant impacts were between equal-sized bodies, rather than one by
one. Furthermore, the majority of the mass accreted is likely due to
collisions close to head on, which deliver a smaller random component of
angular momentum than grazing ones \citep{AA04}. On the other hand, grazing
collisions could deliver spin angular momentum and little mass, allowing
easily twice the naive number of giant impacts. These uncertainties could be
addressed using hydrodynamic simulations. 

\section{ACCRETION AFTER GIANT IMPACTS}
The stirring force protoplanets exert on each other can be balanced by the
force due to dynamical friction caused by the planetesimals as long as $\sigma
> \Sigma$, ensuring small random velocities of the protoplanets. However, as
the protoplanets accrete more planetesimals, their surface density increases
and dynamical friction becomes less and less effective until it is no longer
able to balance the mutual stirring of the protoplanets. Orbit crossing and
giant impacts set in when $\sigma \sim \Sigma$ \citep{GLS042}. Planetesimal
accretion continues and additional 'new' planetesimals are produced as
byproducts of giant impacts. The exact amount of smaller particles produced in
a giant impact depends on the mass ratio of the two colliding protoplanets,
their relative velocity and impact angle. For example, for collisions between
like-sized protoplanets with an impact velocity of twice their escape velocity
and an impact angle of $30^{ \circ}$ (where $0^{ \circ}$ corresponds to a
head-on collision) about 10\% of the total mass of the system escapes as
smaller particles \citep{AA04}. Due to the production of 'new' planetesimals
in giant impacts and the fact that giant impacts set in when $\sigma \sim
\Sigma$, large amounts of planetesimals are expected to still be present after
the culmination of giant impacts. This is also required to relax the high
eccentricities of planets expected after giant impacts \citep{GLS042}. N-body
simulations predict eccentricities of $\sim 0.1$ for terrestrial planets after
giant impacts \citep{C01}. The eccentricity damping timescale $t_{damp}$ due
to dynamical friction caused by left-over planetesimals is given by
\begin{equation}
t_{damp}=-v \frac{dt}{dv} \sim \frac{\rho R}{\sigma \Omega}
\left(\frac{v}{v_{esc}}\right)^4 .
\end{equation}
This timescale should be shorter than the time required for the remaining
planetesimals to be accreted onto the terrestial planets:
\begin{equation}
t_{acc}=-\sigma \frac{dt}{d\sigma} \sim \frac{\rho R}{\Sigma
  \Omega}\left(\frac{v}{v_{esc}}\right)^2.
\end{equation}
This yields
\begin{equation}
\sigma \gtrsim \Sigma \left(\frac{v}{v_{esc}}\right)^2 \sim 0.07 \Sigma.
\end{equation}
Therefore, more than 7\% of the mass should still reside in planetesimals in
order to damp the planets' eccentricities. The eventual accretion of the
remaining planetesimals delivers additional angular momentum to the
planet. For sufficiently small planetesimals, this accretion would again be in
the semi-collisional or collisional regime and hence deliver additional
prograde angular momentum to the planet. The accretion of about 10\%
$M_\earth$ in semi-collisional manner would be sufficient to deliver an
angular momentum equivalent to that of the Earth-Moon system. For Mars less
than 3\% of its mass would need to be accreted semi-collisionally to supply
its current angular momentum, assuming that it had no previous spin. These
small percentages indicate that semi-collisional or collisional accretion of
only a small fraction of the planet's mass after giant impacts is sufficient
to substantially alter planetary spins leading again to favoritism of prograde
rotation. Formation of gaps in the planetesimal disk after giant impacts may
complicate this picture.

\section{CONCLUSIONS}
We have shown here, that planetesimal accretion might be in the
semi-collisional or collisional regime leading to the formation of
a prograde accretion disk around the protoplanet. Such a disk gives rise to
a maximally-spinning protoplanet with prograde rotation. The final
spin of terrestrial planets is therefore no longer random but is biased toward
prograde rotation. The dominance of prograde rotation might be increased
further by the accretion of leftover planetesimals in the post-giant impact
phase, provided that semi-collisional or collisional accretion still
applies. Comparing our results with the spin properties of the terrestrial
planets is somewhat difficult since the spins of Mercury and Venus have
evolved considerably since their formation \citep{LR93,MD93}, leaving only
Earth and Mars whose spins have evolved to a much lesser degree. Earth and
Mars both display prograde rotation with small obliquities, which is
consistent with semi-collisional or collisional accretion. However, no firm
conclusions can be drawn from such a small data set and we cannot rule out
that the low obliquities of Earth and Mars are coincidental. 

Terrestrial planet formation in the asteroid belt was interrupted when growing
planets became massive enough to gravitationally perturb the local population,
causing bodies to collide with increased energy, ending accretion and
commencing fragmentation. Evidence from Vesta's crust \citep[e.g.][]{C86} and
recent models of collisional evolution in the astroid belt \citep{GH97,B05}
suggest that the largest asteroids have survived un-shattered and that they
experienced very little collisional evolution. Their current spin properties
may therefore still contain some information about their primordial spin state
and hence clues about the formation of protoplanets \citep{D89,B05}. The two
most massive asteroids, Ceres and Vesta, both exercise prograde rotation with
periods of 9.1 and 5.3 hours respectively. Ceres' spin axis has a $12^{
\circ}$ inclination with respect to the normal of the ecliptic \citep{T05} and
Vesta's spin axis inclination to the normal of the ecliptic is $\sim 40^{
\circ}$ \citep{DFC98}. The spin properties of Ceres and Vesta might therefore
be indicative of semi-collisional or collisional accretion in the asteroid
belt.

Kuiper Belt objects (KBOs) grew mainly by planetesimal accretion. The
formation time for Pluto-sized KBOs is comparable to the time required for a
collisional cascade to set in, grinding initially kilometer-sized
planetesimals to meters in size. If indeed a collisional cascade started by
the time the largest KBOs formed, semi-collisional accretion could have
dominated their formation. This may explain the intriguingly rapid spin of
$\rm{ 2003~ EL_{61}}$, whose rotation period is only $\sim 4~{\rm hours}$
\citep{R06}. However, the retrograde rotations of Pluto and $\rm{2003~
EL_{61}}$ \citep{BB05} (assuming that it spins in the same direction as it is
orbited by its largest satellite) conflict with this and tentatively suggest
that semi-collisional accretion did not dominate their formation.

\clearpage

\end{document}